\documentclass[preprint2]{aastex62}

\received{August 18, 2017}
\revised{2017}
\accepted{\today}
\submitjournal{ApJ}

\shorttitle{A new classical nova shell around a CV}
\shortauthors{Guerrero et al.}

\begin{document}

\title{Discovery of a new classical nova shell around a nova-like
cataclysmic variable}

\correspondingauthor{Mart\'\i n A. Guerrero}
\email{mar@iaa.es}

\author[0000-0002-7759-106X]{Mart\'\i n A. Guerrero}
\affil{Instituto de Astrof\'\i sica de Andaluc\'\i a, IAA-CSIC, 
Glorieta de la Astronom\'\i a s/n, 18008 Granada, Spain}

\author{Laurence Sabin}
\affiliation{Instituto de Astronom\'\i a, Universidad Nacional Aut\'onoma 
de M\'exico, Apdo.\ Postal 877, C.P.\ 22860, Ensenada, B.C., Mexico}

\author{Gagik Tovmassian}
\affiliation{Instituto de Astronom\'\i a, Universidad Nacional Aut\'onoma 
de M\'exico, Apdo.\ Postal 877, C.P.\ 22860, Ensenada, B.C., Mexico}

\author{Edgar Santamar\'\i a}
\affiliation{Centro Universitario de Ciencias Exactas e Ingenier\'\i as, 
CUCEI, Universidad de Guadalajara, Av.\ Revoluci\'on 1500, C.P.\ 44860, 
Guadalajara, Jalisco, Mexico}
\affiliation{Instituto de Astronom\'\i a y Meteorolog\'\i a, Dpto.\ de 
F\'\i sica, CUCEI, Universidad de Guadalajara, Av.\ Vallarta 2602, C.P.\ 
44130, Guadalajara, Jalisco, Mexico}

\author{Raul Michel}
\affiliation{Instituto de Astronom\'\i a, Universidad Nacional Aut\'onoma 
de M\'exico, Apdo.\ Postal 877, C.P.\ 22860, Ensenada, B.C., Mexico}

\author{Gerardo Ramos-Larios}
\affiliation{Instituto de Astronom\'\i a y Meteorolog\'\i a, Dpto.\ de 
F\'\i sica, CUCEI, Universidad de Guadalajara, Av.\ Vallarta 2602, C.P.\ 
44130, Guadalajara, Jalisco, Mexico}

\author{Alexandre Alarie}
\affiliation{Instituto de Astronom\'\i a, Universidad Nacional Aut\'onoma 
de M\'exico, Apdo.\ Postal 877, C.P.\ 22860, Ensenada, B.C., Mexico}

\author{Christophe Morisset}
\affiliation{Instituto de Astronom\'\i a, Universidad Nacional Aut\'onoma 
de M\'exico, Apdo.\ Postal 877, C.P.\ 22860, Ensenada, B.C., Mexico}

\author{Luis C. Berm\'udez Bustamante}
\affiliation{Instituto de Astronom\'\i a, Universidad Nacional Aut\'onoma 
de M\'exico, Apdo.\ Postal 877, C.P.\ 22860, Ensenada, B.C., Mexico}

\author{Chantal P. Gonz\'alez}
\affiliation{Instituto de Astronom\'\i a, Universidad Nacional Aut\'onoma 
de M\'exico, Apdo.\ Postal 877, C.P.\ 22860, Ensenada, B.C., Mexico}

\author{Nick J. Wright}
\affiliation{Astrophysics Group, Keele University, Keele ST5 5BG, UK}



\begin{abstract}

The morphology and optical spectrum of IPHAS\,XJ210205+471015, a nebula 
classified as a possible planetary nebula, are however strikingly similar 
to those of AT\,Cnc, a classical nova shell around a dwarf nova.  
To investigate its true nature, we have obtained high-resolution
narrow-band [O~{\sc iii}] and [N~{\sc ii}] images and deep GTC 
OSIRIS optical spectra.
The nebula shows an arc of [N~{\sc ii}]-bright knots notably enriched
in nitrogen, whilst an [O~{\sc iii}]-bright bow-shock is progressing
throughout the ISM.  
Diagnostic line ratios indicate that shocks are associated
with the arc and bow-shock. 
The central star of this nebula has been identified by its photometric 
variability. 
Time-resolved photometric and spectroscopic data of this source reveal 
a period of 4.26 hours which is attributed to a binary system. 
The optical spectrum is notably similar to that of RW\,Sex, a cataclysmic 
variable star (CV) of the UX\,UMa nova-like (NL) type.  
Based on these results, we propose that IPHASX\,J210205+471015 is a 
classical nova shell observed around a CV-NL system in quiescence.

\end{abstract}




\section{Introduction} \label{sec:intro}

Many Galactic emission line surveys have been completed in recent years.    
Most of them have focused on the H$\alpha$ line \citep[e.g., the Southern
H$\alpha$ Sky Survey Atlas SHASSA, the INT Photometric H$\alpha$ Survey
IPHAS, the SuperCOSMOS H$\alpha$ Survey SHS, and the VST Photometric H$\alpha$
Survey of the Southern Galactic Plane and Bulge
VPHAS+,][]{Gaustad_etal01,Drew_etal05,Parker_etal05,Drew_etal14}, but also on
molecular emission lines of H$_2$ \citep[e.g., the UKIRT Widefield Infrared
Survey for H$_2$ UWISH,][]{Froebrich_etal11}.  
These surveys offer the possibility to detect different
types of diffuse emission sources and emitting-line stars
\citep{Witham_etal08,Froebrich_etal15}.  
Among the former, many new 
planetary nebulae \citep[PNe,][]{Viironen_etal09,Sabin_etal14},
Wolf-Rayet (WR) nebulae \citep{SB10,Gvaramadze_etal10},
novae \citep{Sahman_etal15,Wesson_etal2008}, 
nebulae ejected from massive evolved stars \citep{Wright_etal2014}, and
supernova remnants \citep{Sabin_etal13} candidates have been uncovered, 
whereas among the latter many emission line star candidates such as
chromospherically active T-Tauri and late-type stars
\citep{Barentsen_etal11,Kalari_etal15},
Be stars \citep{Raddi_etal13}, massive early stars \citep{MS_etal15},
interacting binaries and cataclysmic variables \citep{PK08}, and symbiotic stars
\citep[SS,][]{Corradi_etal10} have been reported.

Follow-up spectroscopic observations are in most cases required
to verify the nature of these sources.  
This has been the case of Galactic PNe, whose number has increased notoriously
\citep[see the compilations in the Macquarie and Hong-Kong/Australian
Astronomical Observatory/Strasbourg Observatory H$\alpha$ Planetary
Nebula MASH PN and HASH PN databases, ][]{Parker_etal06,PBF16}.  
In an attempt to confirm the nature of IPHAS PN candidates,
\citet{Sabin_etal14} has carried out a spectroscopic follow-up of
IPHAS sources and identified 159 new true, likely, and possible
PNe.
Among these sources, the nebula IPHASX\,J210204.7$+$471015 (aka
PN\,G088.0$+$00.4, hereafter J210204), classified as possible PN,
has very peculiar morphology and spectral properties.  
The IPHAS H$\alpha$+[N~{\sc ii}] image reveals an arc-like of emission
which is reminiscent of the morphology seen in the nova shell around
the dwarf nova (DN) AT\,Cnc \citep{Shara_etal12}.
Moreover, the optical spectrum reveals a lack of Balmer lines which is not
typical for PNe, but frequently seen in nova shells \citep{Tarasova14}.  
The morphology and spectral properties of J210204 cast doubts on its
PN nature and rather point to a nova event at its origin.

Nebular shells around novae are scarce \citep{Sahman_etal15}, but they
reveal interesting information to date the nova event, to study the
details of the nova ejection, and to investigate its interaction with
the circumstellar (CSM) and interstellar (ISM) media.
This has motivated us to undertake an observational project to
investigate the real nature of J210204.  
Deep narrow-band optical images have been used to study in detail its
morphology, whereas deep optical spectra of its nebular emission have
been obtained to assess its spectral properties.  
The central star has been identified and photometric and spectroscopic
monitoring have been used to study its variability.
These observations are described in Section \S2.
The nebular morphology and spectral properties, and the stellar
properties are presented in Section \S3.  
The nature and evolutionary status of J210204 are discussed in Section \S4.

\begin{figure*}
\begin{center}
\plottwo{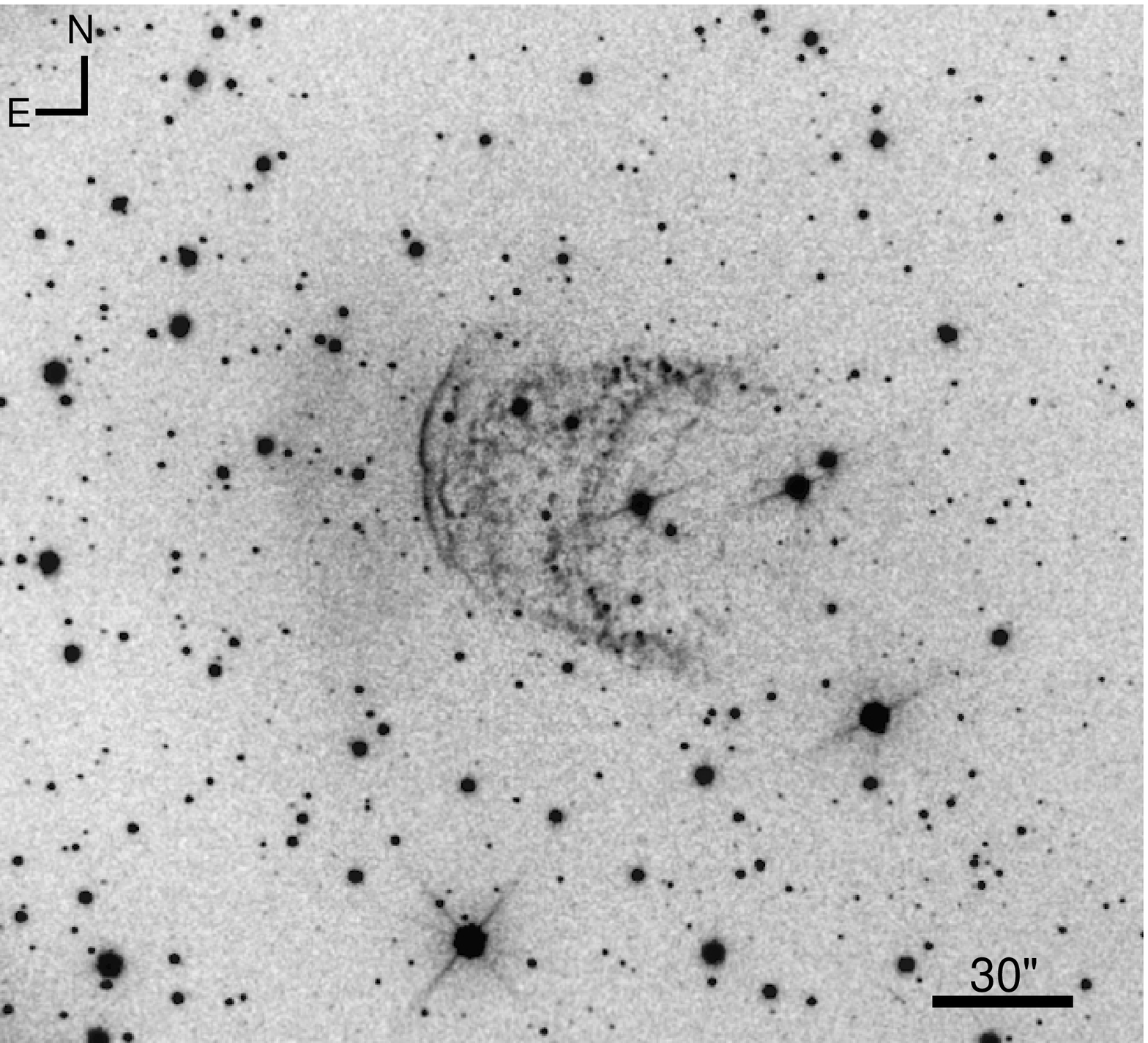}{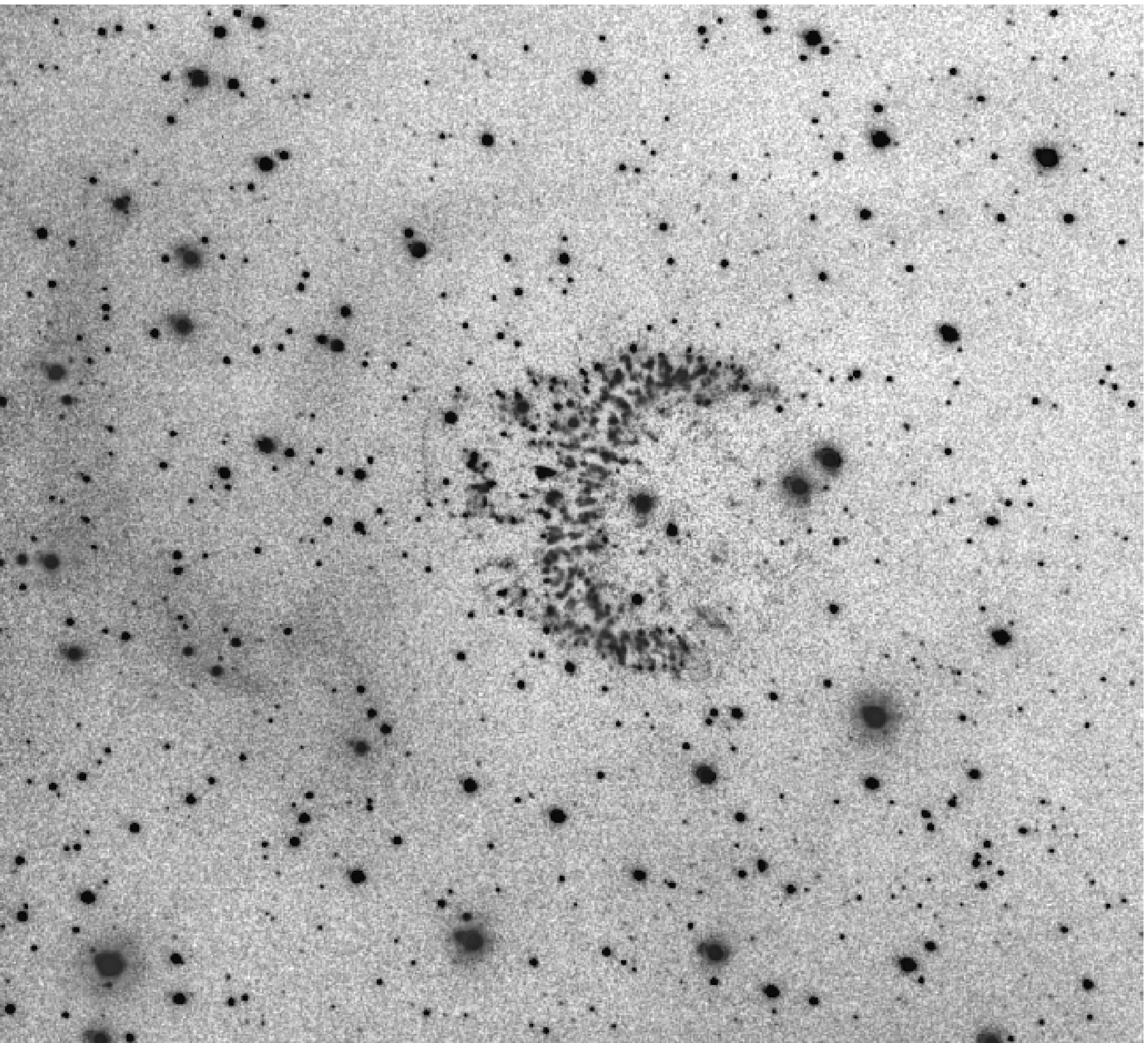}
\caption{
NOT ALFOSC [O~{\sc iii}] (left panel) and [N~{\sc ii}] (right panel)
narrow-band images of J210204.
}
\label{img}
\end{center}
\end{figure*}

\begin{figure*}
\begin{center}
\includegraphics[width=0.970\textwidth]{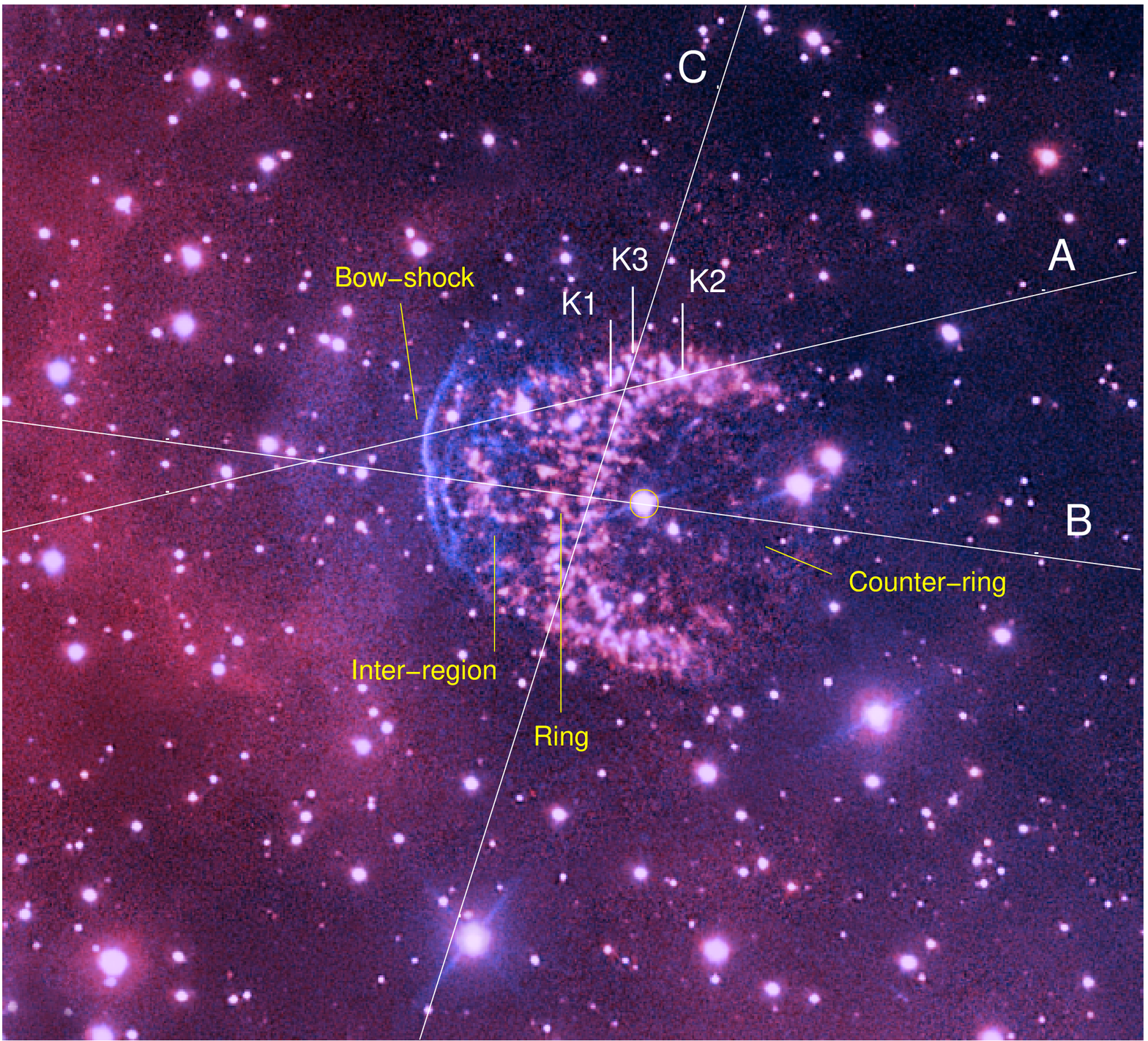}
\caption{
NOT ALFOSC color-composite picture of J210204 using the [N~{\sc ii}] (red) 
and [O~{\sc iii}] (blue) narrow-band images presented in Figure~\ref{img}.
The positions of the GTC OSIRIS long-slits are marked on the picture.  
The different morphological components whose spectra are shown in
Figure~\ref{GTC_spec} are labeled, as well as the location of the
three individual knots K1, K2, and K3 whose spectra are shown in
Figure~\ref{GTC_knots_spec}.
The central star is encircled.  
}
\label{pict}
\end{center}
\end{figure*}

\section{Observations}

\subsection{NOT Optical Imaging}

Narrow-band optical images in the [O~{\sc iii}] $\lambda$5007 and
[N~{\sc ii}] $\lambda$6583 lines were obtained on May 27 and 28,
2017 using the Andalucia Faint Object Spectrograph and Camera
(ALFOSC) mounted at the Cassegrain focus of the 2.5m Nordic Optical
Telescope (NOT) at the Observatorio de El Roque de los Muchachos
(ORM, La Palma, Spain).
The images were obtained using the NOT \#90 [O~{\sc iii}] 501\_3 and 
the Observatorio de Sierra Nevada (OSN) [N~{\sc ii}] \#E16 narrow-band 
filters with central wavelenghts (and full-width half maximum, $FWHM$)
of 5007 \AA\ (30 \AA) and 6584 \AA\ (10 \AA).
The EEV 231--42 2K$\times$2K CCD was used, providing a pixel scale of 
0\farcs211 and a field-of-view (FoV) of 7\farcm0.

Three 1200s and two 900s were secured in the [O~{\sc iii}] and 
[N~{\sc ii}] filters, respectively, to remove cosmic rays.
The images were bias subtracted and flat-fielded using appropriate
sky flat field images, and then combined using {\sc IRAF}\footnote{
  IRAF is distributed by the National Optical Astronomy Observatories,
  which are operated by the Association of Universities for Research
  in Astronomy, Inc., under cooperative agreement with the National
  Science Foundation.
  }
standard routines.
The spatial resolution of the images, as derived from stars in the FoV,
is 0\farcs7.  
The [O~{\sc iii}] and [N~{\sc ii}] images of J210204 are presented
in Figure~\ref{img}, together with a composite color picture in 
Figure~\ref{pict}.

\subsection{SPM  Photometric and Spectroscopic Monitoring}

\subsubsection{Photometry}

The CCD camera Mexman mounted on the 0.84m telescope of the Observatorio
Astron\'omico Nacional at San Pedro M\'artir (OAN SPM, Mexico) was used
to perform time-resolved photometry of the stellar field around J210204.  
The E2V-4240 CCD (``Spectral Instrument 1'') detector was used.
The detector has 2048$\times$2048 pixels with size 13.5 $\mu$m,
but a 2$\times$2 binning was used during the observations.
The resulting plate scale and FoV were 0\farcs434 and
7\farcm4$\times$7\farcm4, respectively.
Alternating \emph{BVRI} observations where taken with exposure times
of 60, 40, 20, and 15-20s, respectively.
A preliminary inspection of these datasets reveals the variability of the
bright star at the geometrical center of J210204 (see Figure~2), which was
then selected as the most likely progenitor of this nebula.  
The coordinates of this star, as implied from the astrometric measurements
and corrections for all IPHAS objects \citep{Barentsen_etal14}, are
$\alpha_{J2000}$ = 21$^{h}$ 02$^{m}$ 05$^{s}$.82,
$\delta_{J2000}$ = +47$^{\circ}$ 10$\arcmin$ 18$\arcsec$.00.   
The positional error of this source, as referenced to the 2MASS data
catalogue, is 0\farcs06.  
For consistency, we will keep the designated IPHAS name for the central
star, which hence is IPHAS\,J210205.83+471018.0.

The log of the photometric observations can be found in Table~\ref{log_phot}.
Data were reduced using standard {\sc IRAF} aperture photometry routines.
In preparation for this work, the same field was observed on
August 26, 2016 and calibrated with Landolt standard stars.  
The star 2MASS\,J21020346+4707113, with measured magnitudes of $B$=14.04,
$V$=13.42, $R$=13.01, and $I$=12.57, was found to be the best reference
star, since it has the most similar colors to those of the target star.

\begin{deluxetable}{lccc}[h!]
\tablecaption{Photometric observing log \label{log_phot}}
\tablecolumns{4}
\tablewidth{0pt}
\tablehead{
\colhead{Date} & \colhead{HJD start time} & \colhead{HJD end time} & 
\colhead{Coverage} \\
\colhead{} & \colhead{+2457600} & \colhead{+2457000} & \colhead{} \\
\colhead{} & \colhead{(day)}    & \colhead{(day)}    & \colhead{(hours)} 
}
\startdata
2016-08-31 & 31.795520 & 31.971839 & 4.2 \\
2016-09-01 & 32.808253 & 32.971219 & 3.9 \\
2016-09-02 & 33.817217 & 33.965623 & 3.6 \\
2016-09-29 & 60.688186 & 60.815632 & 3.1 \\
2016-10-01 & 62.661248 & 62.876114 & 5.2 \\
2016-10-02 & 63.693037 & 63.836978 & 3.4 \\
2016-10-03 & 64.665781 & 64.821321 & 3.7 \\
2016-10-05 & 66.622653 & 66.734812 & 2.7 \\
\enddata
\end{deluxetable}

\subsubsection{Spectroscopy}

Time-resolved spectroscopic data were obtained with the Boller \& Chivens
spectrograph installed on the OAN SPM 2.1m telescope.
We obtained a series of spectra using the 1200 l~mm$^{-1}$ grating,
which provides a spectral dispersion of 1.16 \AA~pix$^{-1}$.
Coupled with the slit-width of 1\farcs5, this results in a
spectral resolution $\simeq$3.2 \AA, suitable to study radial
velocity variations of $\simeq$30 km~s$^{-1}$.  
The 600 l~mm$^{-1}$ grating was also used to obtain low resolution
spectra covering a wide range of wavelengths in the optical domain.
The CuArNe lamp was employed to obtain wavelength calibration arcs.
The spectro-photometric standards Feige\,110, BD+28$^\circ$4211 and G\,191B2B
were observed each night to allow the flux calibration of the source spectra.  
Standard procedures, including bias and flat-field correction,
cosmic ray removal, and wavelength and flux calibration were
applied using IRAF routines.  
Table~\ref{log_spec} presents the log of these spectrophotometric
observations.

\begin{deluxetable*}{ccllrcrrr}[hb!]
\tablecaption{Spectroscopic observing log \label{log_spec}}
\tablecolumns{4}
\tablewidth{0pt}
\tablehead{
\colhead{Date} & 
\colhead{HJD} & 
\colhead{Telescope} & 
\colhead{Instrument} & 
\colhead{Spectral} & 
\colhead{Spectral} & 
\colhead{Number of} & 
\colhead{Exposure} & 
\colhead{Time} \\
\colhead{} & 
\colhead{+2450000} & 
\colhead{} & 
\colhead{} & 
\colhead{coverage} & 
\colhead{dispersion} & 
\colhead{exposures} & 
\colhead{time} & 
\colhead{coverage} \\
\colhead{} & 
\colhead{(day)} & 
\colhead{} & 
\colhead{} & 
\colhead{(\AA)} & 
\colhead{(\AA~pix$^{-1}$)} & 
\colhead{} & 
\colhead{(s)} & 
\colhead{(m)} 
}
\startdata
2016-08-10 & 7610 & OAN 2.1m & B\&Ch  &  4400-5620 & 0.59 &  5~~~~~~ & 1200~~~~ & 100~~~ \\
2016-08-11 & 7611 & OAN 2.1m & B\&Ch  &  4400-5620 & 0.59 &  5~~~~~~ & 1200~~~~ & 100~~~ \\
2016-08-12 & 7612 & OAN 2.1m & B\&Ch  &  4400-5620 & 0.59 &  9~~~~~~ & 1200~~~~ & 200~~~ \\
2016-08-14 & 7614 & OAN 2.1m & B\&Ch  &  4400-5620 & 0.59 &  6~~~~~~ & 1200~~~~ & 120~~~ \\
2016-08-15 & 7615 & OAN 2.1m & B\&Ch  &  4400-5620 & 0.59 &  6~~~~~~ & 1200~~~~ & 120~~~ \\
2016-08-28 & 7628 & GTC      & OSIRIS &  3650-7820 & 2.07 &  8~~~~~~ &  340~~~~ &  75~~~ \\  
2016-08-28 & 7628 & GTC      & OSIRIS & 5110-10340 & 2.59 &  8~~~~~~ &  300~~~~ &  75~~~ \\
2016-09-25 & 7656 & OAN 2.1m & B\&Ch  &  4150-6510 & 1.16 &  9~~~~~~ &  900~~~~ & 140~~~ \\
2016-09-26 & 7657 & OAN 2.1m & B\&Ch  &  4350-6710 & 1.16 & 15~~~~~~ &  900~~~~ & 230~~~ \\
\enddata
\end{deluxetable*}

\subsection{GTC Optical Spectra}

Intermediate-resolution long-slit spectroscopic observations were carried
out with OSIRIS (Optical System for Imaging and low-Intermediate-Resolution
Integrated Spectroscopy) at the 10.4m GTC telescope of the ORM on June 10,
2016 and August 28, 2016.
OSIRIS was used in standard mode, with two Marconi CCD42-82
(2048$\times$4096 pixels) detectors.
The 2$\times$2 on-chip binning resulted in a spatial scale of
0\farcs254~pix$^{-1}$. 
The R1000B and R1000R grisms were used to acquire spectra in the blue and
red regions of the optical spectrum, respectively, covering the spectral
range from 3630 \AA\ to 10000 \AA.  
Their spectral dispersions are 2.12 \AA\ pix$^{-1}$ and 2.62 \AA\ pix$^{-1}$,
respectively.  
A slit-width of 1\farcs2 was used, resulting in a spectral resolution
$\simeq$9 \AA.

The slit, with length 7\farcm4, was placed along different position
angles (PA) to cover distinct morphological components of J210204.  
The details of the observations are listed in Table~\ref{tab_GTC}, including 
the slit keyword (as shown in Figure~\ref{pict}), the grism, PA, offset from 
the central star, and number of exposures and integration time at each slit 
position.
These datasets are also described in Table~\ref{log_spec}.

The data reduction was carried out using standard {\sc IRAF} routines.
Hg-Ar, Ne, and Xe arcs were used for wavelength calibration.  
The spectro-photometric standards G191-B2B and Feige\,110 were used for
flux calibration.  
The seeing, as determined from the FWHM of the continuum of field stars
covered by the slit, was $\simeq$1\farcs0.

\begin{deluxetable}{cccccr}[h!]
\tablecaption{Details of the GTC OSIRIS observations \label{tab_GTC}}
\tablecolumns{6}
\tablewidth{0pt}
\tablehead{
\colhead{Position} & 
\colhead{Grism} & 
\colhead{PA} & 
\colhead{offset} & 
\colhead{N$_{\rm exp}$} & 
\colhead{t$_{\rm exp}$} \\
\colhead{} & 
\colhead{} & 
\colhead{($^\circ$)} & 
\colhead{($^{\prime\prime}$)} & 
\colhead{} & 
\colhead{(s)}
}
\startdata
A & R1000B & 103  & 24.6 & 2 & 738 \\
A & R1000R & 103  & 24.6 & 2 & 600 \\
B & R1000B & 83.5 &   0  & 8 & 340 \\
B & R1000R & 83.5 &   0  & 8 & 300 \\
C & R1000B & 164  & 12.7 & 2 & 738 \\
C & R1000R & 164  & 12.7 & 2 & 600 \\  
\enddata
\end{deluxetable}

\section{Data Analysis}

\subsection{The Nebula}

\subsubsection{Morphology}

The most noticeable structure revealed by the narrow-band images is a
[N~{\sc ii}]-dominated arc-like feature (Figs.~\ref{img}-\emph{right}
and \ref{pict}).  
It consists of a myriad of knots distributed along two parallel strings
that interweave at their tips.
Many of those knots have a cometary morphology, with tails pointing
outwards.
A few knots also trace an Eastern counterpart of the arc, but they are
much fainter and sparse.
The Western double arc feature is also detected in [O~{\sc iii}], but
the emission in this line is much smoother (Fig.~\ref{img}-\emph{left}).
We will refer to these two regions as the ``ring'' and the
``counter-ring'', respectively, as marked in Figure~\ref{pict}.

There is an additional string of knots West of the ring, 
whose emission is similarly bright in the [N~{\sc ii}] and
[O~{\sc iii}] emission lines.  
It must be noted, however, that the [N~{\sc ii}] emission from this
third arc arises from a sparse ensamble of discrete knots, whereas
the [O~{\sc iii}] emission does from a smooth arc.
We will refer to this feature as the ``inter-region'' (Fig.~\ref{pict}).

There is another arc of emission external to the previous ones.
This arc is notably bright in [O~{\sc iii}] with
very faint [N~{\sc ii}] emission.  
Contrary to the previous morphological features described above,
the emission from this double-arc structure is notably sharp.  
Its morphology is clearly reminiscent of a bow-shock structure,
and so we will refer to it as the ``bow-shock'' (Fig.~\ref{pict}).

Along with these main morphological components, there are two additional
arcs of diffuse emission in front of the ``bow-shock'' structure: 
the first one detected both in [O~{\sc iii}] and [N~{\sc ii}]
at $\simeq$1\farcm5 from the central star (Figs.~\ref{img} and
\ref{pict}), the second one detected only in [N~{\sc ii}] at
$\simeq$2\farcm0 from the central star (Figs.~\ref{img}-right
and \ref{pict}).
This second arc along the direction of the [O~{\sc iii}] bow-shock
is confirmed in large-scale IPHAS H$\alpha$+[N~{\sc ii}] images.

\subsubsection{Spectral Properties}

\begin{figure}
\begin{center}
\includegraphics[bb=41 65 355 420,width=0.5\textwidth]{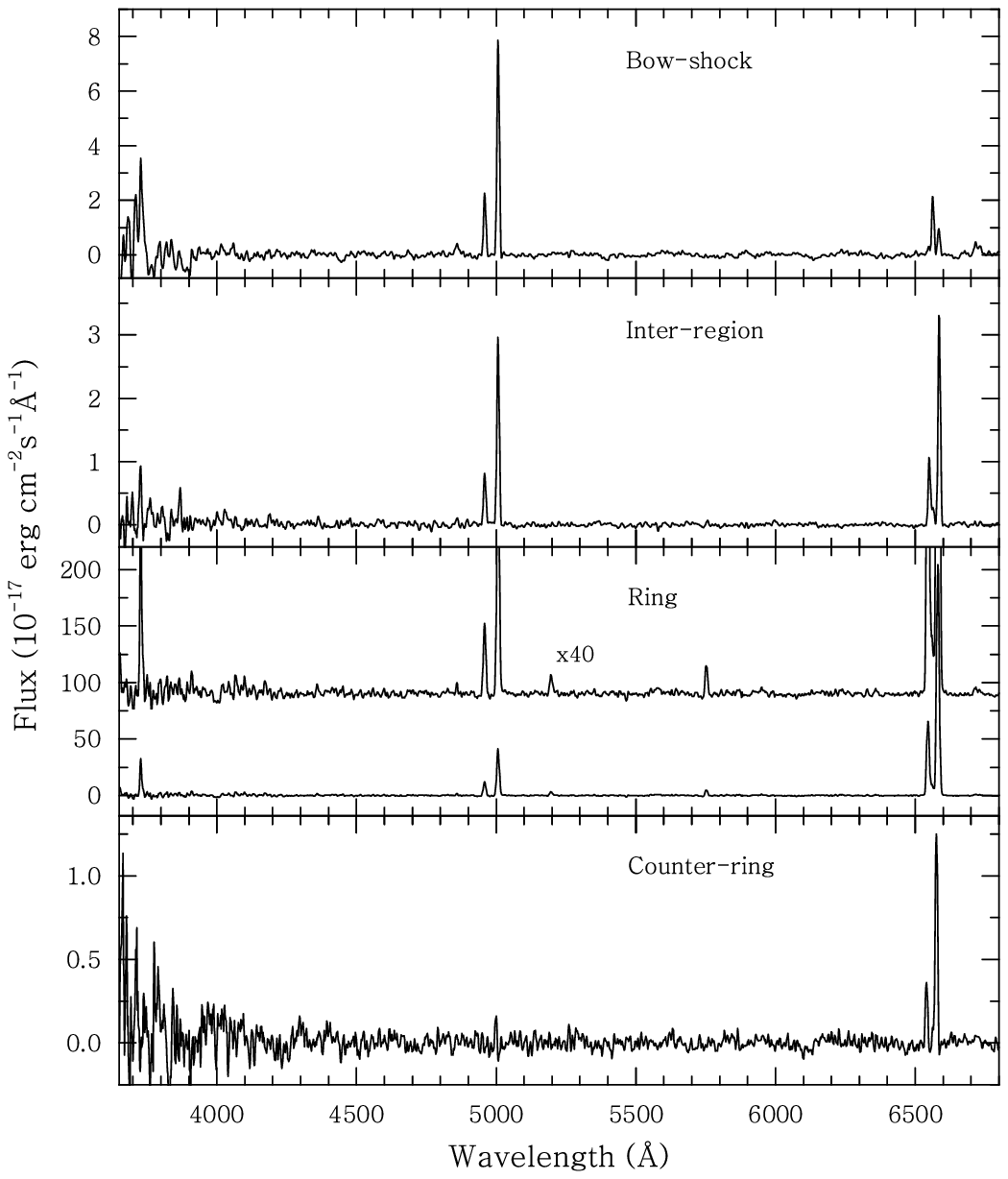}
\caption{
GTC OSIRIS one-dimensional spectra of the different nebular components
of J210204.
The spectrum of the brightest region, the ``Ring'', is plotted at two
different scales to show both the bright and faint emission lines.  
}
\label{GTC_spec}
\end{center}
\end{figure}

We have used the deep GTC OSIRIS observations to extract one-dimensional
spectra of the different morphological components of J210204.
The spectra, shown in Figure~\ref{GTC_spec}, corresponds to
the bow-shock structure, the ring of [N~{\sc ii}]-bright knots,
the inter-region,
and the counter-ring.

As expected, the spectrum of the bright ring is dominated by the
[N~{\sc ii}] $\lambda\lambda$6548,6584 emission lines.
The auroral [N~{\sc ii}] $\lambda$5755 emission line is also
detected, as well as the [N~{\sc i}] $\lambda$5199 emission
line.
The [O~{\sc iii}] $\lambda\lambda$4959,5007 lines and the
[O~{\sc ii}] $\lambda$3727 doublet are also detected, but
they are much fainter than the [N~{\sc ii}] emission lines.
Weak emission from the [S~{\sc ii}] $\lambda\lambda$6717,6731 doublet
is also detected.
The Balmer lines are very weak and no helium lines are detected.

\begin{figure}
\begin{center}
\includegraphics[bb=41 20 355 290,width=0.5\textwidth]{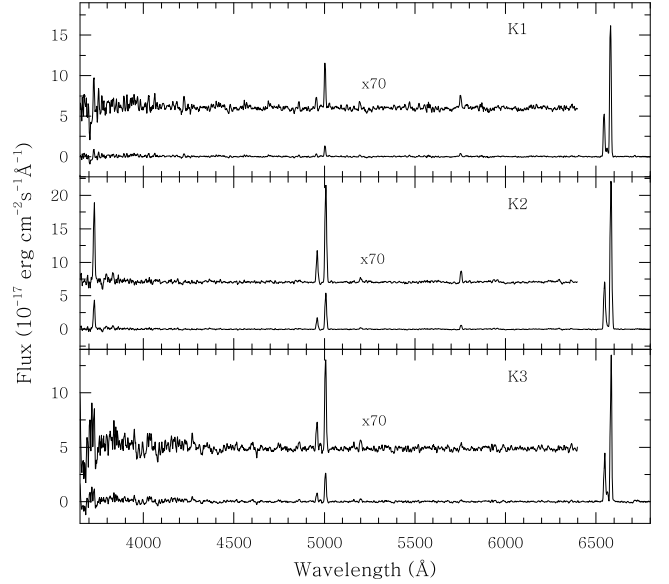}
\caption{
GTC OSIRIS one-dimensional spectra of three selected knots of J210204
as marked in Figure~\ref{pict}.  
The spectra are shown at two different scales to highlight both the
bright and faint emission lines.  
}
\label{GTC_knots_spec}
\end{center}
\end{figure}

We note that the spectrum of different knots may vary significantly
among them.
We have selected three individual knots, marked in
Figure~\ref{pict} as K1, K2, and K3, to illustrate
these variations.  
The spectra of these knots are shown in Figure~\ref{GTC_knots_spec}.
Knot K1 has the lowest [O~{\sc iii}]/[N~{\sc ii}] and
[O~{\sc ii}]/[N~{\sc ii}] line ratios.
Knots K2 and K3 have [O~{\sc iii}]/[N~{\sc ii}] line ratios typical
of the ring, $\sim$0.3, but the [O~{\sc ii}]/[N~{\sc ii}] line ratios
differ notably, from $\sim$0.5 in K2 to $\sim$0.2 in K3.

\begin{deluxetable*}{lccccccc}[h!]
\tablecaption{Emission line flux measured in different regions of J210204 
\label{tab_flux}} 
\tablecolumns{8}
\tablewidth{0pt}
\tablehead{
\colhead{Line ID} & 
\colhead{Bow-shock} & 
\colhead{Inter-region} & 
\colhead{Ring} & 
\colhead{Counter-ring} & 
\colhead{Knot \#1} & 
\colhead{Knot \#2} & 
\colhead{Knot \#3} \\
\colhead{} & 
\colhead{} & 
\colhead{} & 
\colhead{} & 
\colhead{($10^{-16}$ erg cm$^{-2}$ s$^{-1}$)} & 
\colhead{} & 
\colhead{} & 
\colhead{} 
}
\startdata
{[}O~{\sc ii}{]} 3727  &   4.16  &   1.62  &  30.9   & $\dots$ &   1.34  &   4.30  &   0.93  \\
He~{\sc ii} 4686       &   0.20  & $\dots$ & $\dots$ & $\dots$ & $\dots$ & $\dots$ & $\dots$ \\
H$\beta$               &   0.47  &   0.052 &   1.55  & $\dots$ &   0.17  &   0.08  &   0.25  \\
{[}O~{\sc iii}{]} 4959 &   2.53  &   0.93  &  13.8   & $\dots$ &   0.47  &   1.83  &   0.83  \\
{[}O~{\sc iii}{]} 5007 &   8.82  &   3.24  &   4.66  &   0.19  &   1.22  &   5.97  &   2.76  \\
{[}N~{\sc i}{]}   5199 & $\dots$ & $\dots$ &   3.66  & $\dots$ &   0.16  &   0.30  &   0.26  \\  
{[}N~{\sc ii}{]}  5755 & $\dots$ & $\dots$ &   5.09  & $\dots$ &   0.43  &   0.72  &   0.14  \\
{[}N~{\sc ii}{]}  6548 &   0.31  &   1.12  &  81.2   &   0.39  &   6.10  &   8.33  &   4.53  \\
H$\alpha$              &   2.35  &   0.27  &  10.3   & $\dots$ &   1.19  &   0.65  &   0.91  \\
{[}N~{\sc ii}{]}  6584 &   1.05  &   3.55  & 254     &   1.33  &  18.1   &  26.4   &  13.8   \\
{[}S~{\sc ii}{]}  6716 &   0.52  &   0.044 &   1.64  & $\dots$ &   0.18  & $\dots$ & $\dots$ \\ 
{[}S~{\sc ii}{]}  6731 &   0.35  &   0.035 &   1.03  & $\dots$ & $\dots$ & $\dots$ & $\dots$ \\ 
\enddata
\end{deluxetable*}

\begin{deluxetable*}{lccccccc}[h!]
\tablecaption{Dereddened relative intensity of the emission lines in J210204
\label{tab_int}}
\tablecolumns{8}
\tablewidth{0pt}
\tablehead{
\colhead{Line ID} & 
\colhead{Bow-shock} & 
\colhead{Inter-region} & 
\colhead{Ring} & 
\colhead{Counter-ring} & 
\colhead{Knot \#1} & 
\colhead{Knot \#2} & 
\colhead{Knot \#3} 
}
\startdata
 {[}O~{\sc ii}{]} 3727  &  14.6$\pm$5.6  &   51$\pm$13    & 32.9$\pm$1.6   &   $\dots$    & 13.5$\pm$3.5   &   95$\pm$5      &  0.20$\pm$0.08 \\
 He~{\sc ii} 4686       &  0.47$\pm$0.23 &   $\dots$      &   $\dots$      &   $\dots$    &   $\dots$      &    $\dots$      &   $\dots$      \\
 H$\beta$               &  1.00$\pm$0.22 &  1.0$\pm$1.0   &  1.00$\pm$0.18 &   $\dots$    &  1.00$\pm$0.29 &   1.0$\pm$0.8   &  1.00$\pm$0.22 \\
 {[}O~{\sc iii}{]} 4959 &  5.19$\pm$0.20 & 17.0$\pm$0.5   &  8.59$\pm$0.20 &   $\dots$    &  2.65$\pm$0.29 &  23.5$\pm$0.7   &  3.22$\pm$0.22 \\
 {[}O~{\sc iii}{]} 5007 & 17.70$\pm$0.25 & 58.4$\pm$0.5   & 28.51$\pm$0.26 &  22$\pm$7    &  6.90$\pm$0.35 &  75.4$\pm$0.8   & 10.44$\pm$0.25 \\
 {[}N~{\sc i}{]} 5199   &   $\dots$      &   $\dots$      &  2.07$\pm$0.15 &   $\dots$    &  0.88$\pm$0.29 &   3.5$\pm$0.6   &  0.91$\pm$0.19 \\  
 {[}N~{\sc ii}{]} 5755  &   $\dots$      &   $\dots$      &  2.37$\pm$0.12 &   $\dots$    &  1.83$\pm$0.22 &   6.9$\pm$0.5   &  0.41$\pm$0.12 \\
 {[}N~{\sc ii}{]} 6548  &  0.37$\pm$0.11 & 12.27$\pm$0.27 & 30.16$\pm$0.15 & 27.9$\pm$2.4 & 20.87$\pm$0.16 &  63.8$\pm$0.4   & 10.40$\pm$0.11 \\
 H$\alpha$              &  2.86$\pm$0.13 &  3.00$\pm$0.27 &  3.80$\pm$0.08 &   $\dots$    &  4.05$\pm$0.15 &   4.94$\pm$0.32 &  2.09$\pm$0.11 \\
 {[}N~{\sc ii}{]} 6584  &  1.27$\pm$0.13 & 38.46$\pm$0.31 & 93.46$\pm$0.27 & 93.5$\pm$2.6 & 58.82$\pm$0.22 & 200.0$\pm$0.5   & 31.25$\pm$0.16 \\
 {[}S~{\sc ii}{]} 6716  &  0.61$\pm$0.11 &  0.42$\pm$0.19 &  0.58$\pm$0.07 &   $\dots$    &  0.59$\pm$0.14 &    $\dots$      & $\dots$        \\ 
 {[}S~{\sc ii}{]} 6731  &  0.41$\pm$0.11 &  0.35$\pm$0.23 &  0.36$\pm$0.06 &   $\dots$    &    $\dots$     &    $\dots$      & $\dots$        \\ 
\enddata
\end{deluxetable*}

The spectrum of the counter-ring looks very similar to that of the ring,
although it is much fainter.  
On the contrary, the spectrum of the bow-shock feature is
dominated by the [O~{\sc iii}] $\lambda\lambda$5007,4959
emission lines.
H$\alpha$ emission is present, as well as H$\beta$.
Indeed, H$\alpha$ emission is brighter than the [N~{\sc ii}] emission.  
Actually, owing to the presence of large-scale diffuse H$\beta$
and H$\alpha$ emissions, which make difficult the subtraction of
the background emission, this is the only region where the
detection of the Balmer lines are reliable.

Finally, the intensity of the [N~{\sc ii}] $\lambda$6584 and
[O~{\sc iii}] $\lambda$5007 emission lines is very similar in
the inter-region, i.e., their ratio shows an averaged
behaviour between those of the bow-shock and ring spectra.
The H$\alpha$ line is detected, but the intensity of the
H$\beta$ line is rather unreliable.

We show in Table~\ref{tab_flux} the emission line fluxes measured
for the different regions of J210204.
The fluxes of the [N~{\sc ii}] $\lambda\lambda$6548,6584 and
[O~{\sc iii}] $\lambda\lambda$5007,4959 emission lines can be
considered to be reliable, within a few percent, except for
the [O~{\sc iii}] $\lambda$5007 emission line in the counter-ring
region.  
The fluxes of the H$\alpha$ emission line in the bow-shock spectra and
the [O~{\sc ii}] $\lambda$3727 doublet and [N~{\sc ii}] $\lambda$5755
and [N~{\sc i}] $\lambda$5199 emission lines in the ring spectrum are
also reliable.

We have used the H$\alpha$ to H$\beta$ ratio in the bow-shock
region to derive a logarithmic extinction of 0.73.
This correction has been applied to all regions.
The relative intensities in Table~\ref{tab_int} have been referred to that 
of H$\beta$ equal to 1, but for the region of the counter-ring, where the 
Balmer lines are not detected.  
In this region, the relative intensities of the [N~{\sc ii}] 
and [O~{\sc iii}] lines have been referred to that of the 
[N~{\sc ii}] $\lambda$6584 line in the ring.

\subsubsection{Excitation and Chemical Abundances}

The relative line intensities listed in Table~\ref{tab_int} are 
revealing.  
The weak emission in the H$\beta$ and H$\alpha$ lines implies that
all regions have mostly high line ratios of [S~{\sc ii}]/H$\alpha$,
[N~{\sc ii}]/H$\alpha$, [O~{\sc iii}]/H$\beta$, and/or
[O~{\sc ii}]/H$\beta$:  wherever the [S~{\sc ii}] lines are detected, 
the [S~{\sc ii}]/H$\alpha$ line ratio is larger than 0.25, and the
ring has [O~{\sc ii}]/H$\beta$ and [N~{\sc ii}]/H$\alpha$ line
ratios $\simeq$30 and $\simeq$25, respectively.
The spectrum of the bow-shock is remarkably different to that of
the ring, as it exhibits a low [N~{\sc ii}]/H$\alpha$ line ratio
and emission in the He~{\sc ii} $\lambda$4686 line.  
The line ratios detected in the different regions of J210204
discard a photo-ionization excitation, but they rather suggest
that line emission is mostly excited by shocks \citep{HRH87}.

The MAPPINGS 5.13 code \citep{SD17} has been used to investigate key line 
ratios detected in the bow-shock, inter-region and ring to assess the 
properties of these shocks and the chemical abundances of the emitting 
material.
Unfortunately, the available emission line ratios are insufficient 
to fit simultaneously all parameters that critically determine the
plasma emitting properties, mostly the shock velocity, chemical 
abundances, pre-ionization state of the gas, age of the shock, 
pre-shock density, and to a minor extent the magnetic field.

\begin{figure}
\begin{center}
\includegraphics[bb=54 150 558 628,width=0.5\textwidth]{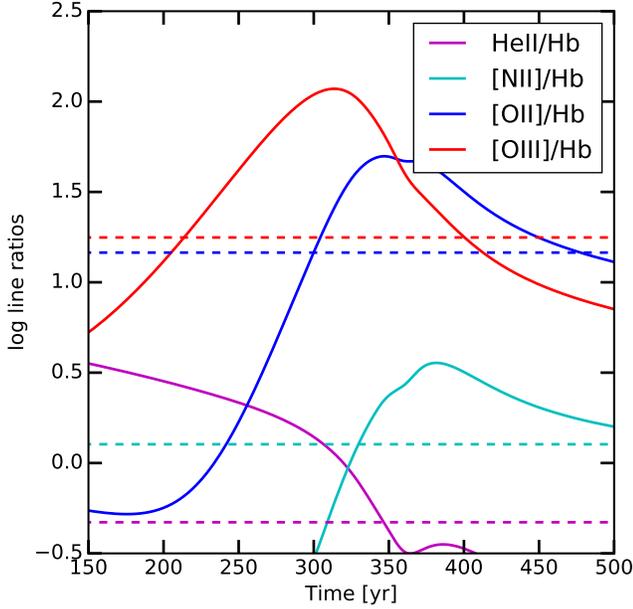}
\caption{
Time evolution of different line ratios for MAPPINGS simulations of an 
incomplete shock at a shock velocity 150 km~s$^{-1}$, pre-shock density 
5~cm$^{-3}$, and solar abundances as suitable for the bow-shock.  
}
\label{MAPPINGS:bow-shock}
\end{center}
\end{figure}

As for the bow-shock, a set of models was built using MAPPINGS 5.13 and 
its line ratio predictions compared to those listed in Table~\ref{tab_int}.  
Since the shocked gas is expected to be dominated by ISM material, its 
chemical abundances have been assumed to be solar \citep{Aetal2009}, 
and the pre-shock density and magnetic field have been adopted to be 
5~cm$^{-3}$ and 1 $\mu$G, respectively.  
The MAPPINGS models show that complete shocks lead to very low values of the 
[O~{\sc iii}]/H$\beta$ line ratios, indicating that the shock is incomplete, 
i.e., the shock has not had enough time to fully recombine and cool down yet.  
The predicted time evolution of some line ratios for a shock 
velocity of 150 km~s$^{-1}$ and the observed values are shown 
in Figure~\ref{MAPPINGS:bow-shock}.  
The He~{\sc ii}/H$\beta$, [O~{\sc ii}]/H$\beta$, [O~{\sc iii}]/H$\beta$, 
and [N~{\sc ii}]/H$\beta$ line ratios are reproduced at times varying 
between 200 and 500 years.  
Line ratios vary on a first-order linearly with the elemental 
abundances, making everything fit at an age $\sim$350 yr if 
the oxygen abundances were reduced by a factor of three.  
We note, however, the strong degeneracy between age, shock velocity and 
abundances and the obvious simplicity of the model, which does not account 
for geometrical effects that superpose shocks of differing properties along 
the line of sight.

\begin{figure}
\begin{center}
\includegraphics[bb=26 185 550 600,width=0.5\textwidth]{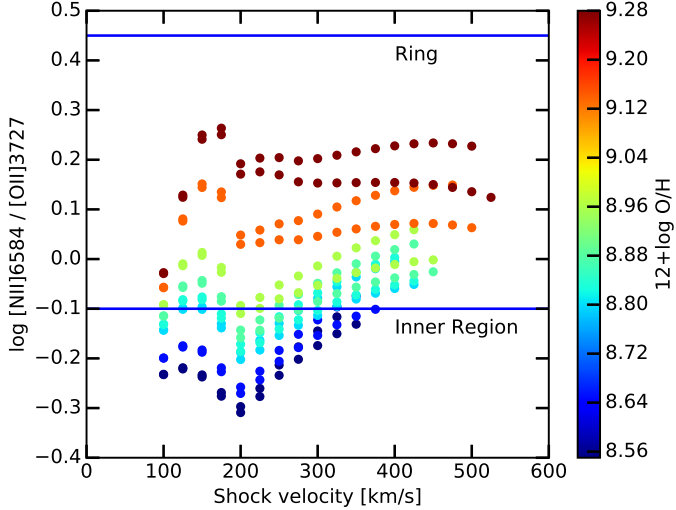}
\caption{
Variation of the [N~{\sc ii}]/[O~{\sc ii}] line ratios for MAPPINGS 
simulations of a complete shock vs.\ the shock velocity and O/H 
abundances for a pre-shock density 1,000~cm$^{-3}$, solar N/O 
abundances ratio, and magnetic field of 1 and 5 $\mu$G.  
}
\label{MAPPINGS:line-ratio}
\end{center}
\end{figure}

As for the ring and inter-region, the significant [N~{\sc ii}]/[O~{\sc ii}] 
line ratio variations between the ring (2.8) and inter-region (0.8) suggest 
the chemical enrichment of the material at the ring.  
The variations of the [N~{\sc ii}]/[O~{\sc ii}] line ratio have been 
simulated using also MAPPINGS for a range of shock velocities and O/H 
abundances, adopting this time a pre-shock density of 1000 cm$^{-3}$, 
consistent with the nova CSM material, and assuming the shock has had 
enough time to become complete.  
The results, plotted in Figure~\ref{MAPPINGS:line-ratio}, are shown for
solar N/O ratios, but the [N~{\sc ii}]/[O~{\sc ii}] line ratio can be
scaled with the N/O ratio.  
This plot reveals that, for shock velocities above 200 km~s$^{-1}$, 
the [N~{\sc ii}]/[O~{\sc ii}] line ratio can be used as a diagnostic 
of the O/H abundances and N/O abundances ratio. 
The high [N~{\sc ii}]/[O~{\sc ii}] line ratio of the ring 
can only be interpreted with extremely high over-solar 
abundances (O/H$>$4$\times$(O/H)$_\odot$) or over-solar 
N/O ratios (N/O$\simeq$5$\times$(N/O)$_\odot$), or both 
of them, whereas the inter-region shows O/H abundances 
or N/O abundances ratio slightly over-solar.  
The large variations in the [N~{\sc ii}]/[O~{\sc ii}] line ratio among
discrete knots in the ring can be similarly interpreted as evidence of
large variations in their metallicity.  

\subsection{The Central Star}

\subsubsection{Spectral Properties}

\begin{figure}
\begin{center}
\includegraphics[bb=18 165 570 710,width=8.5cm]{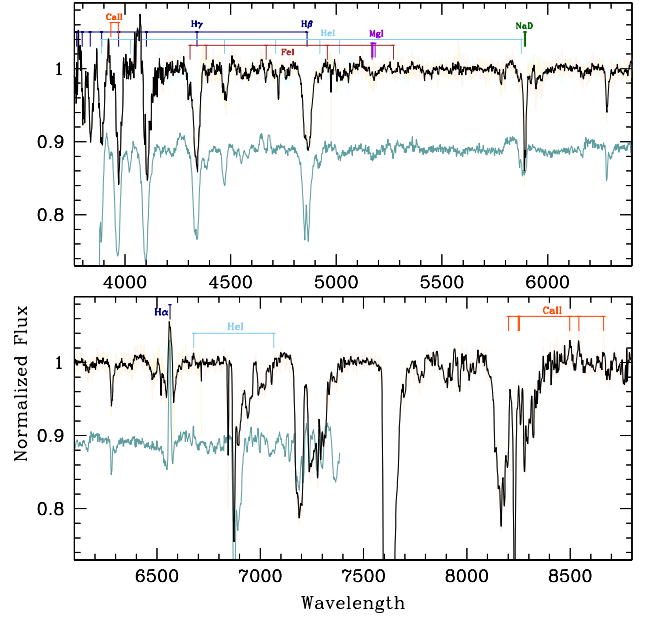}
\caption{
Normalized combined spectrum (SPM+GTC) of the central star of J210204 
(black), compared to that of the UX\,UMa nova-like system RW\,Sex 
(green).  
The spectrum of RW\,Sex has been shifted by 0.11.  
Lines of interest are labeled.  
}
\label{spec}
\end{center}
\end{figure}

The normalized spectrum of the central star of J210204 presented in 
Figure~\ref{spec} reveals prominent broad Balmer hydrogen absorption 
lines with embedded central emission peaks.  
Neutral helium lines are also observed as a combination of
wide absorption and narrow emission.  
Strong NaD and a set of Fraunhofer lines are also detected in absorption in
this spectrum, whilst Ca~{\sc ii} and a blend of C~{\sc iii} and fluorescent
N~{\sc iii} lines at 4640-50 \AA\ are clearly seen in emission.  
The most prominent lines are marked with different colors
in Figure~\ref{spec}.

\subsubsection{Photometric Variability}

\begin{figure}
\includegraphics[width=8.5cm,bb=20 185 600 695]{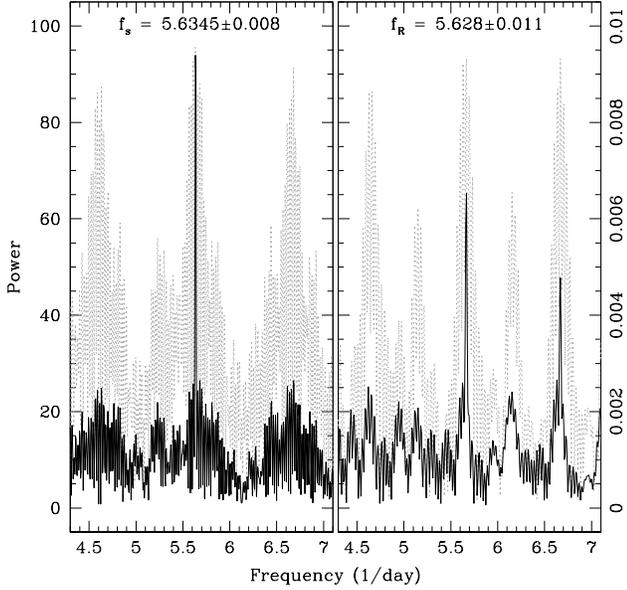}
\hspace{5cm}
\caption{
Power spectra of RV from the H$\beta$ absorption (\emph{left}) and
$R$-band photometric (\emph{right}) time series.
The grey line in both panels is the power calculated by the DFT analysis,
while the solid black line is the same after cleaning (see the text).    
The peak indicates a spectroscopic frequency $f_{s}$=5.6345 day$^{-1}$
and a photometric frequency $f_{R}$=5.628 day$^{-1}$.
}
\label{power}
\end{figure}

\begin{figure}
\hspace{-0.5cm}
\begin{center}
\includegraphics[bb=18 160 592 670,width=8.5cm]{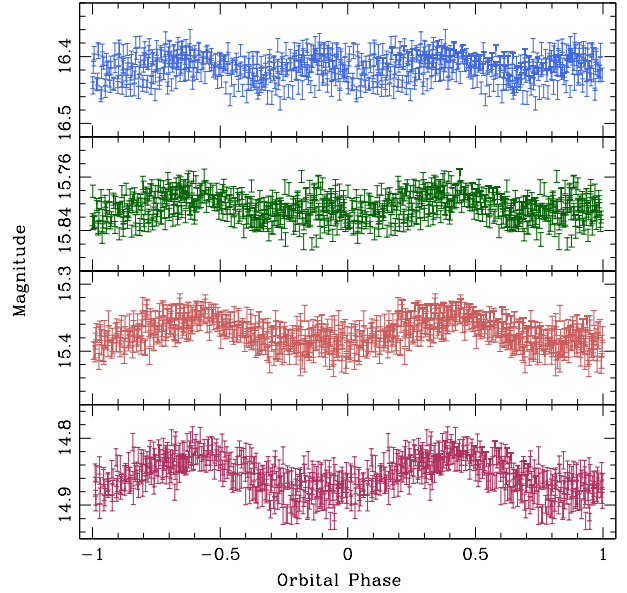}
\caption{
Magnitudes variations folded according to the photometric period of
P$_{ph}=$4.264 hours for the $B$, $V$, $R$, and $I$ bands from top
to bottom, respectively.
}
\label{mag2}
\end{center}
\end{figure}

The measured mean and 1-$\sigma$ deviation magnitudes derived 
from the photometric data obtained during the two main observing
runs are 
16.416$\pm$0.009 mag in the $B$-band,
15.805$\pm$0.009 mag in the $V$-band,
15.373$\pm$0.012 mag in the $R$-band, and
14.856$\pm$0.013 mag in the $I$-band.
The inspection of the light curves  
reveals notable fluctuations.
The discrete Fourier transform (DFT) based program {\sc Period04}
\citep{Lenz2005} was used to search for possible periodicities in
the photometric data.  
The best results showing the lower dispersion are obtained in the $R$-band,
for which a frequency $f_{R}=5.628\pm0.011$ day$^{-1}$ is found as revealed
by the main peak in the power spectrum (dotted grey line in
Figure~\ref{power}-{\it right}).  
One day aliases are strong, given the observational strategy; indeed,
the power spectrum corresponding to the $I$-band actually peaks at
6.63 day$^{-1}$, i.e.\ $f_{I} = f_{R}+1$.
There is also a number of noticeable secondary peaks, resulting from
the significant noise level in the power spectra induced by the low
resolution of the photometric dataset.
The convolution of the observed power with the spectral window helps
to supress aliases using a procedure appropriately called {\sl Clean}
\citep{RLD87}.
The cleaned power spectrum of the $R$-band light curve (solid black
line in Figure~\ref{power}-{\it right}) confirms the low frequency
peak.  
The frequency derived from the $R$-band implies a photometric
period $P_{ph}=$4.264$\pm$0.007 hours.  
The folded light curve with this frequency show little variation
in shape among the different photometric bands (Figure~\ref{mag2}).
The global change in magnitudes throughout the different
bands is relatively small, $\sim$0.1 mag.

There is also a remarkable trend in the long-term light curve.  
The power spectrum reveals the presence of a peak at
$f=0.085 \; {\mathrm {day}^{-1}}$, which would imply
a cyclical variability with a period of 11.7 day,
although we reckon that the length of the observing
runs is not appropriate to investigate suah a long-term
variability.  
Another noteworthy peak is actually a one-day alias at a frequency 
$0.92 \; {\mathrm {day}^{-1}}$ that corresponds to a period of 26.1
hours.
The power strength shifts from one peak to the other
depending on the photometric band.

\subsubsection{Radial Velocity Variability} 

\begin{figure}
\includegraphics[bb=18 165 586 700,width=8.5cm]{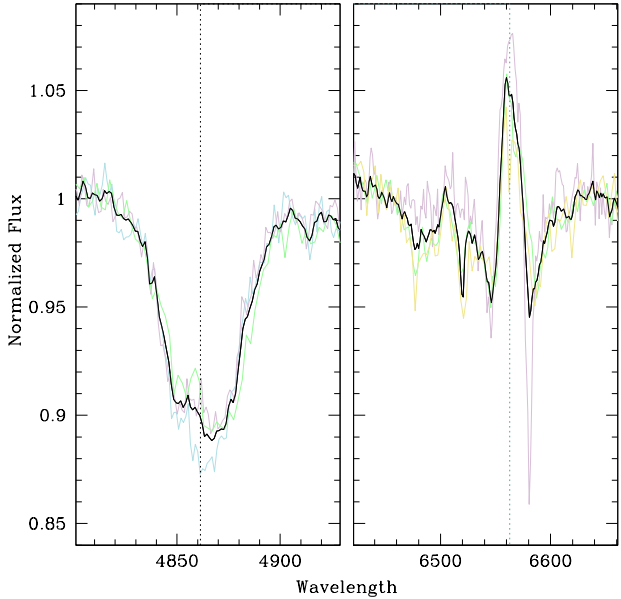}
\caption{H$\beta$ absorption (left) and H$\alpha$ emission (right) profiles.  
The thin colored lines show profiles observed at different times, 
whereas the thick black line correspond to the averaged profiles.  
The dashed vertical lines mark the rest wavelength of both lines.  
}
\label{lines}
\end{figure}

The photometric variability may be associated with radial velocity
(RV) variations.
To investigate these, we have used the narrow emission component of
the Balmer H$\alpha$ line and the Balmer H$\beta$ absorption line 
(see Figure\,\ref{lines}). 
The H$\beta$ line is not symmetric and very wide, with a
$FWHM \approx40-50$ \AA\ and a full-width at zero intensity
($FWZI$) up to $\approx140$ \AA.
In absence of a clear understanding of the nature of  different components
which may form the H$\beta$ line profile, we measured it using a single
Gaussian.
The analysis of the RV variations with {\sc Period04} clearly shows periodic
modulation, with a strong peak around a frequency of 5.6 day$^{-1}$ (with its 
unavoidable one-day alias).
In a similar way as for the photometric light curves, the
uneven time series produces alias peaks in the power spectrum.
These have been convolved with the spectral window to clean aliases.
The raw and cleaned resulting power spectrum are shown in the left
panel of Figure~\ref{power}, where the cleaned power spectrum peak
corresponds to a frequency $f_{s}=5.635\pm0.008$ day$^{-1}$ (or to
a period $P_{s}$= 4.260$\pm$0.006 hours).
This value deviates from the period derived from the photometric data
by $\sim$0.3 minutes, which is within the 3-$\sigma$ formal uncertainty.  
In the following, we will favor the period determined from
spectroscopic observations, as this has been derived from a
more complete time series distribution and the data analysis
shows a more robust variability.  
Additional photometric and spectroscopic data are needed to improve
the accuracy of the period determination to assess whether there is
a real difference between the spectroscopic and the photometric period.

As for the H$\alpha$ line, we measured only the dominant
emission peak located close to the rest wavelength of the
line.
Even though this component looks single-peaked and narrow, compared to
the entire H$\alpha$ complex, it actually splits into two components at
certain orbital phases.
We used the de-blending option in the IRAF procedure
{\sl splot} to fit two distinct Gaussian components
to the line profile whenever they were present.  
The brightest component is basically seen throughout the whole orbital cycle
and it follows the RV curve deduced from H$\beta$, but with a much smaller RV
amplitude.
The RV of the narrower and fainter component, which is detected only
in half of the orbital phases, varies in a near counter-phase to the
first component and to the wide absorption measured with H$\beta$.

Finally, a careful examination of the NaD doublet shows a faint component
with variable RV in addition to the deep narrow absorption lines that can
be attributed to the interstellar and may be circumstellar media.  
The variable component is in phase with the wide H$\beta$ absorption.

\section{On the nature of IPHASX\,J210204.7+471015}

\subsection{The central star IPHAS\,J210205.83+471018.0}

The optical spectrum of the central star of J210204 shows wide absorption
lines of H~{\sc i} and He~{\sc i} with narrow emission cores.
These spectral features are distinctive of UX\,UMa nova-like (NL)
systems \citep{1996ASSL..208....3D}, a sub-class of non-magnetic
cataclysmic variable stars (CVs).
Indeed, the comparison of the optical spectrum of the central star
of J210204 with that of the UX\,UMa-system RW\,Sex
\citep[][and references therein]{Beuermann1992}
obtained with the same settings at the same telescope
\citep{2017arXiv170606236H} show that both spectra are basically
identical (Fig.~\ref{spec}).  
Both stars show spectral lines mostly in absorption, which is
consistent with a rather optically thick accretion disc in a
NL, although sometimes it is difficult to discern them from
similar set of lines produced by a K donor star.
The main difference between the spectrum of RW\,Sex and that of J210204 
central star is the presence of multiple emission peaks inside Balmer 
lines (Figure~\ref{lines}), a fact normally associated with a post-nova 
shell stage. 
Emission lines are limited to the notable H$\alpha$ line, and to the high
excitation lines of He~{\sc ii} 4686 \AA\ and a blend of C~{\sc iii}
and fluorescent {\bf N~{\sc iii}} at $\simeq$4650 \AA.
This places the central star of J210204 on the upper part (or high
state) in the disc surface density ($\Sigma$) vs disc temperature
(T) plot \citep{MO1985,C1993}, i.e., the locus of CV-NL objects, a
sub-class of CVs characterized by high and stable mass transfer
rates ($\dot{M} \geq 10^{-9} M_{\odot}$~yr$^{-1}$) and a steady state
accretion disc \citep{1982A&A...106...34M}. 
Accordingly, the brightness of NLs does not vary significantly around its 
mean level, in agreement with the small $\sim$0.1 mag variation found for 
J210204 (Figure~\ref{mag2}).

The concordant periodic photometric and RV variations of the central star
of J210204 reinforces the idea that it consists of a binary system in a
nova, where the viewing aspect at different orbital phases of the system
comprised by a compact source, an accretion disc around it, and a donor 
star would produce the observed periodic variations.  
The observed 4.26 hour period can be interpreted as the orbital
period of the binary system.  
The short period is again consistent with a CV classification for this source.  
Otherwise, the long-term photometric period of 11.7 days,
although not well understood, is not exceptional in these
systems \citep{2017arXiv170606731Y} and may be caused by
the precession of the accretion disc \citep{2016MNRAS.457.1447D}.

\begin{figure}
\includegraphics[bb=24 164 576 697,width=8.5cm]{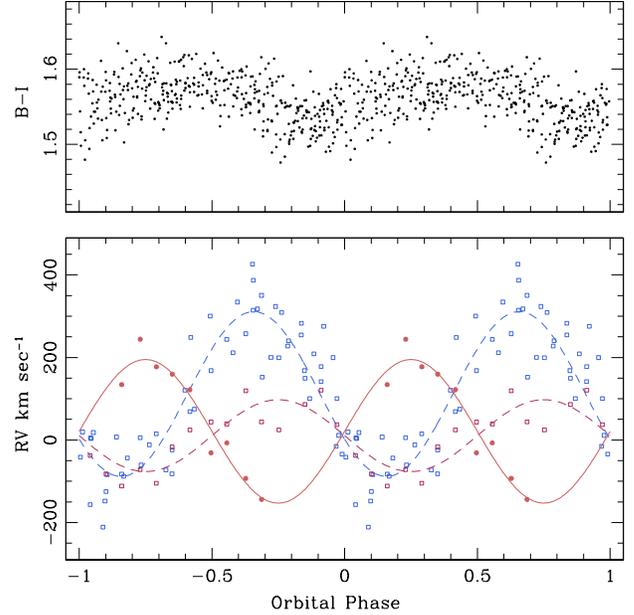}
\caption{
Photometric $B-I$ (top panel) and RV (bottom panel) variability folded with 
the spectroscopic phase.  
In the bottom panel,
the red solid line and dots corresponds to the faint H$\alpha$ component,
the red dashed line and open symbols to the bright H$\alpha$ component, and
the blue dashed line and open symbols to the H$\beta$ absorption. 
}
\label{view}
\end{figure}

In the absence of eclipses, it is difficult to associate the different
emission and absorption line components with the system constituents, 
but the varying amplitudes and phasing of these spectral line components,
as seen in the bottom panel of Figure~\ref{view}, can provide important
clues about this system.
The similarities of the central star of J210204 to other NLs 
\citep{2017arXiv170606236H} makes us assume that the narrow 
faint H$\alpha$ emission component (red solid line and dots 
in Figure~\ref{view}-\emph{bottom}) originates at the irradiated 
face of the donor star.
The RV curve of this component thus reflects the motion
of the secondary star.
Consequently, we will assign the zero phase to the time T0=2457610.5154
at which this RV curve crosses the systematic velocity line.  
When the RV plots are referred to this zero point and folded with
the 4.26 hours spectroscopic period (Figure~\ref{view}-\emph{bottom}),
the emission of the faint H$\alpha$ component apparently fades away
and disappears between phases --0.3 to 0.2.  
This can be taken as a good evidence that this H$\alpha$ emission component 
arises only from the side of the donor star facing the accretion disc that 
would be not visible to the observer in this phase interval. 
The sinusoidal fit to the H$\alpha$ measurements indicates a RV
semi-amplitude of $\sim$175 km~s$^{-1}$.  
The secondary star velocity implies a high orbital 
inclination angle, although not high enough to 
produce eclipses in this system.

Meanwhile, the broad H$\beta$ absorption probably originates in the
optically thick accretion disc.
This optically thick accretion disc would outshine the stellar components
of the binary system, so that no signature of the narrow absorption lines 
from the secondary star is visible.  
If the H$\beta$ absorption line formed evenly in the entire 
accretion disc, then it would be expected its RV curve to be 
in phase with that of the narrow faint H$\alpha$ emission 
component originating from the secondary star, as in the case 
of RW\,Sex \citep[see Figure 6 in][]{Beuermann1992}.
However, the orbital phasing of the H$\beta$ absorption (blue open squares 
and blue dash-line in Figure~\ref{view}-\emph{bottom}) is not exactly in 
counter-phase with the narrow faint H$\alpha$ emission component (red solid 
line and dots in Figure~\ref{view}-\emph{bottom}).  
The maximum positive velocity of the H$\beta$ RV curve at $\phi=0.65$ 
and its large amplitude indicate that the line emanates predominately 
from the co-rotating hot spot.
At that orbital phase, the intrinsic velocity of the matter
in the hot spot would be increased by the orbital velocity.  
This is not a novelty, as the asymmetry of the accretion discs of UX\,UMa 
systems has been noted before \citep[e.g.][]{Schlegel_etal83,Neustroev_etal11}.

Finally, the bright H$\alpha$ emission component (red open
squares and dash-line in the bottom panel of Fig.~\ref{view}) can be
attributed to an extended and low-velocity region located off the edge 
of the accretion disc and opposite to the hot spot, as for 1RXS
J064434.5+334451 \citep{2017arXiv170606236H}.  
This interpretation is consistent with the observed photometric variations 
of the system, as shown in the upper panel of Figure~\ref{view}, where the
$B-I$ color curve folded with spectroscopy ephemeris is plotted.  
The maximum color difference is achieved at $\phi\simeq0.4$, just 
prior to the inferior conjunction of the presumed late type donor 
star. 
The amplitude of the photometric variations grows towards longer wavelengths, 
reaching 0.04-–0.06 magnitude in the $I$-band (Figure\,\ref{mag2}).  
The light curves in red filters are not symmetric, but the $B−I$ color 
curve is, thus suggesting that color stays more or less constant with 
some blue light being obscured around $\phi= 0.9$.  
This might be evidence of the dark spot detected in the archetype UX\,UMa 
itself \citep[see Figure 9 in ][]{Neustroev_etal11}, but our spectrosopic 
data is not sufficient to conduct a similar tomographic analysis and 
confirm the evidence of such dark spot in the central star of J210204.

\subsection{The nebula IPHASX\,J210204.7+471015}

The shell around this binary system bears a striking resemblance 
to the nova AT\,Cnc, particularly in the low-ionization [N~{\sc ii}] 
emission line.  
The cometary morphology of many of the [N~{\sc ii}] knots and the 
[O~{\sc iii}] bow-shock are highly indicative of a recent outburst 
event which is seen as a bow-shock progressing through the ISM.  
Shock-sensitive line ratios are indeed suggestive of shocks 
\citep{HRH87}. 
A detailed spatio-kinematical study of the different nebular components
is underway (Santamar\' ia et al., in preparation).



Novae are expected to produce noticeable amounts of CNO, as well as 
Ne, Mg, Al, and Si \citep[e.g.,][]{Jose1998}, a theoretical 
expectation confirmed by spectroscopic observations of classical 
nova shells \citep{Gehrz2014,Tarasova2016}.
The chemical abundances of the different regions of J210204 cannot be derived 
in detail, but the [N~{\sc ii}]/[O~{\sc ii}] line ratios, in the asssumption of
complete shocks, imply a significant gradient in the abundances from the ring
to the inter-region.
The high N/O ratio of the ring, well above unity as typically 
found in novae \citep{Netal1988}, reveals pristine nova ejecta, 
whereas the inter-region displays chemical abundances closer
to those of the ISM, as the nova ejecta expand and sweep up ISM
material.  
Individual knots may show evidence of larger chemical enrichment and
notably varying N/O ratios, suggesting that material highly enriched
ejected during the outburst has not yet mixed completely.  
The assumption on the shock completeness does not hold at the
bow-shock, where abundances are uncertain, although they might
be consistent with those of the ISM.

\begin{figure}
\begin{center}
\includegraphics[bb=100 0 1112 1185,clip,width=0.475\textwidth]{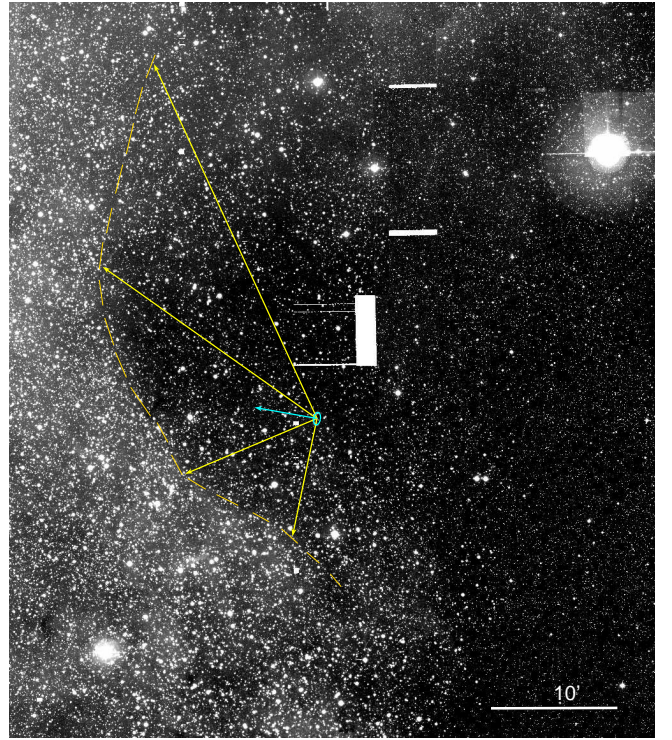}
\caption{
Large field of view IPHAS H$\alpha$+[N~{\sc ii}] image around J210204.
The cyan ellipse traces the bright [N~{\sc ii}] ring of J210204 and the
cyan arrow the direction of the [O~{\sc iii}] bow-shock.
The yellow arrows and dashed line mark the location of the quasi
circular arc described in the text.  
}
\label{FoV}
\end{center}
\end{figure}

The bow-shock is preceded by two additional arcs.
These diffuse arcs of emission seem to trace the interaction of the ejecta of
J210204 with the ISM, although they may also be the relics of previous events
of mass ejection.
In order to investigate recurrent interactions of mass loss processes
from J210224 with the ISM, we have searched for large-scale diffuse
emission features around J210204 in IPHAS mosaic images (Figure~\ref{FoV}).
These do not reveal any additional detached rings or arcs, but
there is a quasi semi circular depression in the interstellar
material towards the north-east of J210204, moreless along the
bow-shock direction (Figure~\ref{FoV}).
Although the presence of this structure is tantalizing, we note that the stars
density inside this arc is notably smaller than on the arc, indicating that
the origin of this arc can be associated most likely with varying interstellar
absorption rather than with recurrent episodes of mass ejection and their
interaction with the ISM.

\section{Conclusions}

From the analysis of the time-resolved photometric and spectroscopic data of 
the central source of IPHASX\,J210204.7+471015, it can be concluded that it 
hosts a binary system.
The object can be classified as an old nova observed around a UX\,UMa 
system, i.e.\ a classical nova observed around a non-magnetic CV system 
observed at quiescence.  
The nebula shows evidences of nitrogen and oxygen enrichment, which
is later diluted into the surrounding ISM by a fast ejecta resulting
in a typical bow-shock structure.  
These are typical characteristics of a nova shell.
Actually, the remnants of past ejections around CVs are scarcely
detected \citep{Ringwald1996, Gill2000, Tomov2015, Schmidtobreick2015}.
A re-classification as a recurrent nova might be possible if the arc-like
features 1\farcm5 and 2\farcm0 in front of the [O~{\sc iii}] bow-shock
could be attributed to previous episodes of mass ejection from the nova.


\acknowledgments

MAG acknowledges support of the grant AYA\,2014-57280-P, cofunded with 
FEDER funds. 
LS acknowledges support from PAPIIT grant IA-101316 (Mexico). 
GR-L acknowledge support from Universidad de Guadalajara, CONACyT, PRODEP 
and SEP (Mexico). 
AA postdoctoral grant and some computational ressources are from 
CONACyT 2015-CB/254132 and UNAM-DGAPA-PAPIIT-107215 projects.
NJW acknowledges an STFC Ernest Rutherford Fellowship.

This article is partially based upon observations carried out at the 
Observatorio Astron\'omico Nacional on the Sierra San Pedro M\'artir 
(OAN-SPM), Baja California, México.
We thank the daytime and night support staff at the OAN-SPM
for facilitating and helping obtain our observations. 
Some of the data presented here were obtained with ALFOSC, which is 
provided by the Instituto de Astrof\'\i sica de Andaluc\'\i a (IAA) 
under a joint agreement with the University of Copenhagen and NOTSA.
The article is based in part on observations made with the Gran
Telescopio Canarias (GTC), instaled in the Spanish Observatorio
del Roque de los Muchachos of the Instituto de Astrof\'\i sica 
de Canarias, in the island of La Palma. 
This paper also makes use of data obtained as part of the INT 
Photometric H$\alpha$ Survey of the Northern Galactic Plane 
(IPHAS, http://www.iphas.org) carried out at the Isaac Newton 
Telescope (INT). 
The INT is operated on the island of La Palma by the Isaac Newton Group 
in the Spanish Observatorio del Roque de los Muchachos of the Instituto 
de Astrof\'\i sica de Canarias. 
All IPHAS data are processed by the Cambridge Astronomical
Survey Unit, at the Institute of Astronomy in Cambridge.
The band-merged DR2 catalogue was assembled at the Centre for
Astrophysics Research, University of Hertfordshire, supported by
STFC grant ST/J001333/1.

%

\vspace{5mm}
\facilities{
ORM:GTC(OSIRIS), 
ORM:INT(WFC), 
ORM:NOT(ALFOSC), 
SPM:0.84m (Mexman), 
SPM:2.1m (B\&Ch)
}


\begin{thebibliography}{}

\bibitem[Asplund et al.(2009)]{Aetal2009}
Asplund, M., Grevesse, N., Sauval, A.~J., \& Scott, P.\ 2009,
\araa, 47, 481 

\bibitem[\protect\citeauthoryear{Barentsen et al.}{2011}]{Barentsen_etal11}
  Barentsen G., et al., 2011,
  MNRAS, 415, 103 

\bibitem[Barentsen et al.(2014)]{Barentsen_etal14} 
Barentsen, G., Farnhill, H.~J., Drew, J.~E., et al.\ 2014, \mnras, 444, 3230 

   \bibitem[Beuermann et al.(1992)]{Beuermann1992} 
 Beuermann, K., Stasiewski, U., \& Schwope, A.~D.\ 1992, \aap, 256, 433 

\bibitem[\protect\citeauthoryear{Cannizzo}{1993}]{C1993} 
  Cannizzo J.~K., 1993, adcs.book, 6 

\bibitem[\protect\citeauthoryear{Corradi et al.}{2010}]{Corradi_etal10}
  Corradi R.~L.~M., et al., 2010,
  A\&A, 509, A41 
  
  \bibitem[\protect\citeauthoryear{de Miguel et al.}{2016}]{2016MNRAS.457.1447D} de Miguel E., et al., 2016, MNRAS, 457, 1447 

 
 \bibitem[\protect\citeauthoryear{Dhillon}{1996}]{1996ASSL..208....3D} Dhillon V.~S., 1996, ASSL, 208, 3

\bibitem[\protect\citeauthoryear{Drew et al.}{2005}]{Drew_etal05}
  Drew J.~E., et al., 2005,
  MNRAS, 362, 753 

\bibitem[\protect\citeauthoryear{Drew et al.}{2014}]{Drew_etal14}
  Drew J.~E., et al., 2014,
  MNRAS, 440, 2036 

\bibitem[\protect\citeauthoryear{Froebrich et al.}{2011}]{Froebrich_etal11}
  Froebrich D., et al., 2011,
  MNRAS, 413, 480 

\bibitem[\protect\citeauthoryear{Froebrich et al.}{2015}]{Froebrich_etal15}
  Froebrich D., et al., 2015,
  MNRAS, 454, 2586 

\bibitem[\protect\citeauthoryear{Gaustad et al.}{2001}]{Gaustad_etal01}
  Gaustad J.~E., McCullough P.~R., Rosing W., Van Buren D., 2001,
  PASP, 113, 1326 

 \bibitem[Gill \& O'Brien(2000)]{Gill2000} 
 Gill, C.~D., \& O'Brien, T.~J.\ 2000, \mnras, 314, 175 

\bibitem[Gehrz et al.(2014)]{Gehrz2014}
  Gehrz, R.~D., Evans, A., \& Woodward, C.~E.\ 2014,
  Stellar Novae: Past and Future Decades, 490, 227 

\bibitem[\protect\citeauthoryear{Gvaramadze et al.}{2010}]{Gvaramadze_etal10}
  Gvaramadze V.~V., Kniazev A.~Y., Hamann W.-R., Berdnikov L.~N., Fabrika S.,
  Valeev A.~F., 2010,
  MNRAS, 403, 760 

\bibitem[\protect\citeauthoryear{Hartigan, Raymond, \& Hartmann}{1987}]{HRH87}
  Hartigan P., Raymond J., Hartmann L., 1987, ApJ, 316, 323 



\bibitem[\protect\citeauthoryear{Hern\'andez et al.}{2017}]{2017arXiv170606236H}
  Hern\'andez M.~S., Zharikov S., Neustroev V., Tovmassian G., 2017, arXiv,
  arXiv:1706.06236

\bibitem[Jos{\'e} \& Hernanz(1998)]{Jose1998}
  Jos{\'e}, J., \& Hernanz, M.\ 1998,
  \apj, 494, 680 

\bibitem[\protect\citeauthoryear{Kalari et al.}{2015}]{Kalari_etal15}
  Kalari V.~M., et al., 2015,
  MNRAS, 453, 1026 


\bibitem[Lenz \& Breger(2005)]{Lenz2005} 
Lenz, P., \& Breger, M.\ 2005, Communications in Asteroseismology, 146, 53 

\bibitem[Linnell et al.(2010)]{Linnell2010} 
Linnell, A.~P., Godon, P., Hubeny, I., Sion, E.~M., \& Szkody, P.\ 2010, 
 \apj, 719, 271 
 
\bibitem[\protect\citeauthoryear{Meyer \& Meyer-Hofmeister}{1982}]{1982A&A...106...34M} 
  Meyer F., Meyer-Hofmeister E., 1982, A\&A, 106, 34

\bibitem[\protect\citeauthoryear{Mineshige \& Osaki}{1985}]{MO1985} 
  Mineshige S., Osaki Y., 1985, PASJ, 37, 1

\bibitem[\protect\citeauthoryear{Mohr-Smith et al.}{2015}]{MS_etal15}
  Mohr-Smith M., et al., 2015,
  MNRAS, 450, 3855 
  
\bibitem[\protect\citeauthoryear{Neustroev et al.}{2011}]{Neustroev_etal11} 
  Neustroev V.~V., Suleimanov V.~F., Borisov N.~V., Belyakov K.~V., Shearer A., 2011, MNRAS, 410, 963 

\bibitem[Nussbaumer et al.(1988)]{Netal1988}
Nussbaumer, H., Schmid, H.~M., Vogel, M., \& Schild, H.\ 1988,
\aap, 198, 179 

\bibitem[\protect\citeauthoryear{Parker et al.}{2005}]{Parker_etal05}
  Parker Q.~A., et al., 2005,
  MNRAS, 362, 689

\bibitem[\protect\citeauthoryear{Parker et al.}{2006}]{Parker_etal06}
  Parker Q.~A., et al., 2006,
  MNRAS, 373, 79 

\bibitem[\protect\citeauthoryear{Parker, Boji{\v c}i\'c, \& Frew}{2016}]{PBF16}
  Parker Q.~A., Boji{\v c}i{\'c} I.~S., Frew D.~J., 2016,
  JPhCS, 728, 032008 

\bibitem[\protect\citeauthoryear{Pretorius \& Knigge}{2008}]{PK08}
  Pretorius M.~L., Knigge C., 2008,
  MNRAS, 385, 1485 

\bibitem[\protect\citeauthoryear{Raddi et al.}{2013}]{Raddi_etal13}
  Raddi R., et al., 2013,
  MNRAS, 430, 2169 

  \bibitem[Ringwald et al.(1996)]{Ringwald1996} 
Ringwald, F.~A., Naylor, T., \& Mukai, K.\ 1996, \mnras, 281, 192 

\bibitem[Roberts et al.(1987)]{RLD87} 
Roberts, D.~H., Lehar, J., \& Dreher, J.~W.\ 1987, 
\aj, 93, 968 

\bibitem[\protect\citeauthoryear{Sabin et al.}{2013}]{Sabin_etal13}
  Sabin L., et al., 2013,
  MNRAS, 431, 279 

\bibitem[\protect\citeauthoryear{Sabin et al.}{2014}]{Sabin_etal14}
  Sabin L., et al., 2014,
  MNRAS, 443, 3388 

\bibitem[\protect\citeauthoryear{Sahman et al.}{2015}]{Sahman_etal15}
Sahman D.~I., Dhillon V.~S., Knigge C., Marsh T.~R., 2015,
\mnras, 451, 2863 

\bibitem[\protect\citeauthoryear{Schlegel, Honeycutt, \& Kaitchuck}{1983}]{Schlegel_etal83} 
Schlegel E.~M., Honeycutt R.~K., Kaitchuck R.~H., 1983,
\apjs, 53, 397 

\bibitem[Schmidtobreick et al.(2015)]{Schmidtobreick2015}
Schmidtobreick, L., Shara, M., Tappert, C., Bayo, A., \& Ederoclite, A.\ 2015,
\mnras, 449, 2215 

\bibitem[\protect\citeauthoryear{Shara et al.}{2012}]{Shara_etal12}
  Shara M.~M., Mizusawa T., Wehinger P., Zurek D., Martin C.~D., Neill J.~D.,
  Forster K., Seibert M., 2012,
  ApJ, 758, 121

\bibitem[\protect\citeauthoryear{Stock \& Barlow}{2010}]{SB10}
Stock D.~J., Barlow M.~J., 2010,
MNRAS, 409, 1429 

\bibitem[Sutherland \& Dopita(2017)]{SD17}
Sutherland, R.~S., \& Dopita, M.~A.\ 2017,
\apjs, 229, 34 

\bibitem[Tarasova(2016)]{Tarasova2016}
Tarasova, T.~N.\ 2016,
Astronomy Reports, 60, 1052 

\bibitem[\protect\citeauthoryear{Tarasova}{2014}]{Tarasova14}
Tarasova T.~N., 2014,
Astronomy Reports, 58, 302 

\bibitem[\protect\citeauthoryear{Tovmassian et al.}{2014}]{2014AJ....147...68T} 
Tovmassian G., Stephania Hernandez M., Gonz{\'a}lez-Buitrago D., Zharikov S.,
Garc{\'{\i}}a-D{\'{\i}}az M.~T., 2014,
\aj, 147, 68

\bibitem[Tomov et al.(2015)]{Tomov2015} 
 Tomov, T., Swierczynski, E., Mikolajewski, M., \& Ilkiewicz, K.\ 2015, 
 \aap, 576, A119 

\bibitem[\protect\citeauthoryear{Viironen et al.}{2009}]{Viironen_etal09}
  Viironen K., et al., 2009,
  A\&A, 504, 291 

\bibitem[\protect\citeauthoryear{Warner}{1995}]{Warner1995} 
Warner B., 1995, CAS, 28,  

\bibitem[Wesson et al.(2008)]{Wesson_etal2008} Wesson, R., Barlow, M.~J., Corradi, R.~L.~M., et al.\ 2008, \apjl, 688, L21 

\bibitem[\protect\citeauthoryear{Witham et al.}{2008}]{Witham_etal08}
  Witham A.~R., Knigge C., Drew J.~E., Greimel R., Steeghs D.,
  G{\"a}nsicke B.~T., Groot P.~J., Mampaso A., 2008,
  MNRAS, 384, 1277 
  
\bibitem[Wright et al.(2014)]{Wright_etal2014} Wright, N.~J., Wesson, R., Drew, J.~E., et al.\ 2014, \mnras, 437, L1 


  \bibitem[\protect\citeauthoryear{Yang et al.}{2017}]{2017arXiv170606731Y} Yang M.~T.-C., et al., 2017, arXiv, arXiv:1706.06731 

\end{thebibliography}
\end{document}